\documentclass[12pt]{article}
\newcounter{num}
\begin{document}
\vspace{1.cm}
\begin{center}
\Large\bf CP violation in semi--leptonic decays of the top quark within MSSM
\end{center}
\vspace{0.5cm}
\begin{center}
{ Xiao-Jun Bi and Yuan-Ben Dai}\\\vspace{3mm}
{\it Institute of Theoretical Physics,
 Academia Sinica, \\ P.O.Box 2735, Beijing 100080, P. R. China }
\end{center}

\vspace{1.5cm}
\begin{abstract}
We calculate the CP--violating effects in the top quark semi--leptonic 
three body decays induced by the supersymmetric
CP--odd phase of the top squark trilinear soft breaking term $\arg(A_t)$.
The light top squark mass is assumed to be close to the top quark 
mass $m_{\tilde{t}}\sim m_t$. The CP--conserving phase is provided 
by the $\chi^+$ and $\chi^0$ cut.
We find that the partial rate asymmetry is in the
0.1\% level. In the most favorable parameter region the decay rate asymmetry
can reach up to 0.55\%. 
\end{abstract}
%{\large PACS number: 12.39.Hg, 13.25.Hw, 13.25.Ft, 12.38.Lg}
%
\section {Introduction}
Top quark physics is sensitive to new physics, 
which may exist near the electro--weak scale, due to its large mass.
Experimental and theoretical research about CP violation in top sector is
one way to reveal new physics.  To study the top quark
CP--odd effects has its own advantage that the uncertainties coming from 
hardron matrix elements can be avoided.

If $m_t$ is near to $m_{new physics}$, the CP--assymmetry effects in top 
quark decays can be induced
by new particles. Until now a lot of works about CP--asymmetry effects
in the top quark decays\cite{review} 
have been done within the supersymmetric model. Most of
those works have made the assumption that the mass of the light top
squark is much smaller than that of the top quark.
 B.Grzadkowski and W.Y.Keung
calculated the CP violating effects induced by
$\tilde{t}\tilde{b}\tilde{g}$-loop\cite{bwy}. This contribution requires the 
condition $m_{\tilde{t}} + m_{\tilde{g}} <  m_t$, which has already been
excluded. E.Christova and M.Fabbrichesi computed the effects induced by
$\tilde{t}\tilde{b}\chi^0$-loop\cite{em} which requires
$m_{\tilde{t}} + m_{\chi^0} <  m_t$. S. Bar--Shalom {\it et al} gave 
the CP asymmetry in top quark decay induced by $\tilde{t}\chi^{+}\chi^0$.
It is at best 0.3\% when light stop mass is between $\sim 50GeV$ and 
$\sim 70GeV$\cite{soni}.

However, if the light top squark mass $m_{\tilde{t}_1}$ is approximately
as heavy as the top quark,
CP--asymmetry effect in top quark
two body decay induced by supersymmetric CP--odd phase will not be
observable because the top squark can not run on shell to produce
the necessary absorptive cut.
In this work we considered this case 
under the assumption that the light chargino is much lighter than
$m_{\tilde{t_1}}$. Under this condition, 
the $\chi^0$( which is always assumed to be the LSP ) and
$\chi^+$ can provide necessary absorptive cut in top quark
three body decays, such as in the process 
$t\rightarrow b\nu_{\tau}\tau$ considered in the work. These two particles can be 
on shell in the top quark three body decay loop diagrams when the
invariant mass of the lepton pair is sufficiently large. 
To our knowledge,
a study on CP asymmetry in the top quark three body decays is missed in
literature.

In the present work the mass of the light top squark is assumed to be above 140GeV.
Taking into account the direct experimental limit on super particles\cite{pdg}
 and the indirect limit coming from neutron EDM limit\cite{wsm},
we take $\mu$ to be real and  
, the lightest neutralino 
to be above
30GeV and the lighter chargino to be above 65GeV. The large mass of
stop leads to relatively small CP violating effects. Nevertheless,
the CP--odd effects can reach up to $0.55\%$ in the most favorable parameter
space. 

The paper is organized as following: in Sec. 2 we analyze the possible new CP 
violating sources in MSSM and present our simplifying assumptions 
in calculation. In sec. 3 we sketch the main steps of
 our calculations. In sec. 4 we present our numerical results and
in sec. 5 and sec. 6 we discuss and summarize our results.
The mass matrices for charginos, neutralinos
and squarks are given in appendix A. In appendix B we give the relevant pieces
of the Lagrangian for our calculations and some analytic results
are presented in appendix C.

\section {CP violating phases in the low energy supersymmtric model}
The most general form of the low energy Lagrangian of MSSM\cite{Haber,j},
which is $SU(3)\times SU(2)\times U(1)$ gauge invariant and does not violate
the SM conservation laws, can be written as
\begin{equation}
\label{1}
{\cal L} = \ \mbox{kinetic terms} + 
\int d^2\theta W + {\cal L}_{\rm soft}\ \ .
\end{equation}
The superpotential $W$ is given by
\begin{equation}
W =\ \epsilon_{ij}(  \mu\hat{H}^1_i\hat{H}^2_j +
l^{IJ} \hat{H}^1_i \hat{L}^I_j \hat{R}^J +
u^{IJ} \hat{H}^2_i \hat{Q}^I_j \hat{U}^J
+d^{IJ} \hat{H}^1_i \hat{Q}^I_j \hat{D}^J )
\end{equation}
where $\epsilon_{12}$ =-1. The hat ` \^{} ' indicates that the corresponding 
letter represents a
superfield. The capital indices I,J denote generations. i,j refer to
the components of a SU(2) doublet. The l,d,u are the Yukawa coupling matrices.
The soft breaking terms can be divided into three pieces,
\begin{equation}
{\cal L}_{\rm soft} = {\cal L}_{\rm scalar}+{\cal L}_{\rm gaugino}
+{\cal L}_{\rm trilinear}\ \ .
\end{equation}
These are the scalar particle mass terms, gaugino mass terms 
and the trilinear
soft breaking terms respectively. They are given by

\begin{eqnarray}
{\cal L}_{\rm scalar}  &=& \epsilon_{ij} \mu_s H^1_iH^2_j -m_{H_1}^2 |H^1_i|^2 -m_{H_2}^2 
|H^2_i|^2 - (m_L^2)^{IJ} {L_i^I}^\ast\cdot L_i^J - \nonumber \\ 
&& (m_R^2)^{IJ} {R^I}^\ast\cdot R^J- 
(m_Q^2)^{IJ} {Q_i^I}^\ast\cdot Q_i^J- \nonumber \\
&& (m_D^2)^{IJ} {D^I}^\ast\cdot D^J - (m_U^2)^{IJ} {U^I}^\ast\cdot U^J \ \ ,
\end{eqnarray}
\begin{eqnarray}
\label{gau}
{\cal L}_{\rm gaugino} = \frac{1}{2} 
({m}_1 \lambda_B \lambda_B +
{m}_2 \lambda_W^i \lambda_W^i + 
{m}_3 \lambda_G^a \lambda_G^a) + H.C.\ \ ,
\end{eqnarray}
\begin{eqnarray}
\label{10}
{\cal L}_{\rm trilinear} &=& \epsilon_{ij} 
(\  (A_E l)^{IJ} H^1_i L^I_j R^J +
 (A_D d)^{IJ} H^1_i Q^I_j D^J +\nonumber \\
&& (A_U u)^{IJ} H^2_i Q^I_j U^I \ ) + H.C.\ \ ,
\end{eqnarray}
where ${m}_1$,${m}_2$,${m}_3$ are U(1), SU(2) and SU(3) 
gauginos masses respectively. Fields in ${\cal L}_{\rm scalar}$ and 
${\cal L}_{\rm trilinear}$ are scalar components of the corresponding
superfields.

In general, all the coupling parameters in the above expressions 
except those of the diagonal terms in ${\cal L}_{\rm scalar}$ may be complex
which may be the CP violating sources. However, not all of them are physical
and, even the physical parameters are too many to be disposed of.
In actual calculations, simplifying assumptions must be made. We get 
the physical CP violating phases by the following steps.

First,
we take the GUT assumption that the ${m_i}$s are universal
at the GUT scale and can be set real by a phase rotation\cite{21}.
Thus, the $m_i$s are real at any scale. Second, we adjust the global phase 
between
the two Higgs superfields so that $\mu_s$ is real. This adjustment 
makes the two vacuum
expectation values of the neutral Higgs fields $v_1$, $v_2$ real\cite{j}.
After the 
adjustment the phases of the two Higgs superfields are fixed and $\mu$ is
complex in general. Third,  $l^{IJ}$, $d^{IJ}$,
and $u^{IJ}$ in the superpotential are diagonalized and the unphysical phases 
are absorbed by quark superfields similar to that done in the 
standard model. This leaves
a CP violating phases $\delta_{KM}$ in the \mbox{Kinetic terms} after 
the superfield are redefined. 
Fourth, to suppress the FCNC process in the SUSY extension of SM, as an
approximation 
we require that all the matrices in ${\cal L}_{\rm soft}$, $m^2_L$,
$m^2_R$, $m^2_Q$, $m^2_U$, $m^2_D$, $A_U$, $A_D$ and $A_E$ are flavor
diagonal in the basis where $l^{IJ}$, $d^{IJ}$ and $u^{IJ}$ are diagonal
(flavor alignment\cite{37}). Then the 
Hermitian matrices in ${\cal L}_{\rm scalar}$ are now all real . 
The phases of all squarks are fixed after the third step.
Therefore, $A_D$, $A_U$ are generally complex. In conclusion, $\delta_{KM}$,
$\arg(\mu)$, $\arg(A_D)$s and $\arg(A_U)$s
 are the CP violating phases in the low energy 
supersymmtric model under our assumptions.

$\arg(\mu)$ can not be larger than
the order $\sim {\cal O} (10^{-2}-10^{-3})$ by the constraint 
from the experimental limit of the neutron EDM\cite{wsm}.
In our calculation, we always take $\mu$ to be real and thus no CP--violating
effects are induced by $\mu$.
The CP violating effects induced
by $A_{D,U}$ are greatly suppressed because they are proportional to the
masses of the corresponding quarks which can be neglected compared with 
the squark mass parameters in ${\cal L}_{\rm scalar}$
except that induced by $A_t$ which is associated with the top 
quark(see the form of squark mass matrices in (A.8) and (A.9) in
Appendix A). Thus, $\arg(A_{t})$ is the only new
CP violating source in our calculation.

After the interaction terms in the potential are diagonalized, the MSSM
 Lagrangian will be
expressed by mass eigenstates instead of gauge eigenstates. The CP violating
phases are then transferred to the gauge interaction vertices
(see Appendix B for related Lagrangian pieces).
This is reflected
by mixing matrices in the interaction vertices.
The mixing matrices $Z^+$,$Z^-$,$Z_N$ which diagonalize
charginos and neutralinos are real if $\mu$ is taken real. The 
mixing matrices $Z_t$ for the top squark
is in general complex due to the complexity of $A_t$.
This implies that the CP violating effects comes from $\arg(A_t)$. The 
mixing matrices will be discussed in detail in the appendix A.

\section {Calculation}
We now discuss the CP violating effects in the process
$t\rightarrow b \nu_{\tau} \bar{\tau}$, which corresponds to diagrams in 
Fig.1, within the MSSM framework. 
First denote the invariant mass of $\bar{\tau}$ and $\nu_\tau$ as 
$\sqrt{q^2}$, where $q = p_{\nu_{\tau}}+p_{\bar{\tau}}$.
$p_{\nu_{\tau}}$
and $ p_{\bar{\tau}} $ are the four--momenta of $\nu_{\tau}$
and $\bar{\tau}$.
We calculated the CP asymmetry when $q^2 > m_W^2$. This condition opens a new
window so that $\chi^+$ and $\chi^0$ cut may give an absorptive part to
the amplitudes
for loop diagrams as depicted in Fig.2.

Several points should be indicated at the moment.\\
(1). There should be a minus sign in front of the amplitudes for box diagrams
relative to that for triangle diagrams. \footnote{
This can be explained by
that we can get the amplitude for Fig.2e from that for Fig.2a 
by interchanging the field operators of $\chi^-$ and
$\bar{\tau}$ at two ends of the $\tilde{\tau}$ propagator and simultaneously
changing $\tilde{\tau}$ to W boson in the Wick contraction procedure.
The interchange of two fermion
operators gives the minus sign. An analogous case on QED can be found in
\cite{Brj}.}
\\(2). There are two kinds of box diagrams. One of them, Fig.2a and 
2b, is fermion number nonconservative\cite{Haber}. \\
(3). We have assumed that the absorptive part of the amplitude for 
the loop diagrams are induced
by $\chi^0$ $\chi$ cut\cite{aws}. $\tilde{t}$ or $\tilde{\tau}$ can also be on 
shell and give new contribution to absorptive part 
under the condition $m_{\tilde{t}}+m_{\chi^0} < m_t$ or 
$m_{\chi} > m_{\tilde{\tau}}$, respectively.
We excluded these two cases for simplicity for the following reasons.
Under our assumption about the stop mass, $m_{\tilde{t}}+m_{\chi^0} < m_t$
can be satisfied only in very narrow SUSY parameter region which 
is simply excluded
in  our calculation. $m_{\chi} > m_{\tilde{\tau}}$ means 
the CP--odd effect will appear as
$\sqrt{q^{2}} > 160 GeV$ (we take $m_{\tilde{\nu}}=130 GeV$) which has too small 
branch ratio and can be ignored.

Two quantities are defined to represent the CP--asymmetry effects,
\begin{eqnarray}
\label{def}
A_{CP}^{t,e}& =&\frac{\Gamma-\bar{\Gamma}}{\Gamma+\bar{\Gamma}}\nonumber\\ 
&=&\frac{
\int_{L_{t,e}}^{m^2_t}dq^2\frac{d\Gamma(q^2)}{dq^2}-
\int_{L_{t,e}}^{m^2_t}dq^2\frac{d\bar{\Gamma}(q^2)}{dq^2}
}{
\int_{L_{t,e}}^{m^2_t}dq^2\frac{d\Gamma(q^2)}{dq^2}+
\int_{L_{t,e}}^{m^2_t}dq^2\frac{d\bar{\Gamma}(q^2)}{dq^2}
}\ \ \ ,
\end{eqnarray}
\begin{equation}
L_t=(\ m_{\chi}+m_{\chi^0}\ )^2,
\end{equation}
\begin{equation}
L_e=(\ 100GeV\ )^2,
\end{equation}
where $\frac{d\Gamma(q^2)}{dq^2}$ and $\frac{d\bar{\Gamma}(q^2)}{dq^2}$
are differential widths of top quark and top anti--quark. $A_{CP}^t$ reflects
the theoretical CP--odd effect appearing when the invariant mass $\sqrt{q^2}$ 
is above the threshold.
$A_{CP}^e$
reflects the experimental CP--asymmetry effect when we
measure the decay events with $q^2 > L_e$, which is fixed to be $(100GeV)^2$
in the work.

We have only considered the contribution to the denominator in Eq.~\ref{def} 
from the tree--level diagrams which gives $\Gamma=\bar\Gamma$. The nominator
comes from the interference of the one--loop diagrams with the 
tree--level diagram, as
\begin{eqnarray}
\Delta\Gamma\ &=&\ \Gamma - \bar{\Gamma}\ =\ \Delta|M|^{2}\cdot \mbox{phase
space}\nonumber \\
&=&(\ |M|^{2}-|\bar{M}|^{2}\ )\cdot \mbox{phase space}\ \ .
\end{eqnarray}
The three body final state ``phase space'' is
\begin{eqnarray}
\mbox{phase space}&=&\ \frac{1}{2m_{t}}\int
\frac{
d^{3}p_{b}
}{
(2\pi)^{3}2E_{b}
}
\frac{
d^{3}p_{\bar{\tau}}
}{
(2\pi)^{3}2E_{\bar{\tau}}
}
\frac{
d^{3}p_{\nu_{\tau}}
}{
(2\pi)^{3}2E_{\nu_{\tau}}
}\nonumber \\
&&\cdot (2\pi)^{4}\delta (p_{t}-p_{b}-p_{\bar{\tau}}-p_{\nu_{\tau}}).
\end{eqnarray}
The above expression multiplied by a 
$\delta((p_{\bar{\tau}}+p_{\nu_{\tau}})^{2}
-q^{2})$ gives the phase space for fixed $q^2$.
$M$, $\bar{M}$ are the amplitudes for the process 
$t\rightarrow b \nu_{\tau} \bar{\tau}$ and its CP conjugate process
$\bar{t}\rightarrow \bar{b} \bar{\nu_{\tau}} \tau$, respectively.

$M$ can be expressed as
\begin{equation}\label{amp} M\ =\ aA_{1}\ +\ \sum_i b^i A_{2}^i,\end{equation}
where the two terms come from tree--level and one--loop diagrams respectively.
a, $b^i$ contain the CP--violating phases both from KM matrix element $V_{tb}$
and from stop mixing matrix elements $Z^{ij}_{t}$.
$A_{2}^i$ develops an absorptive part for $q^2$ beyond 
the threshold. $\bar{M}$ can be expressed as
\begin{equation}\bar{M}\ =\ a^{\ast}A_1 + \sum_i {b^i}^{\ast}A_2^i.
\end{equation}
So, we have
\begin{equation}
\label{aa}
\Delta|M|^2= -4\ \sum_i Im(a^{\ast}b^i) Im(A_{2}^iA_1^{\ast}).
\end{equation}

$ImA_2^i$ is given by Cutkosky rule,
\begin{equation}
\label{cut}
\sum_i b^i ImA_{2}^i=\frac{1}{2}\int d\Phi\ 
\hat{A}(t\rightarrow b\chi^{+}\chi^{0})\ \ 
\hat{A}(\chi^{+}\chi^{0}\rightarrow \bar{\tau}\nu_{\tau})\ ,
\end{equation}
where 
\begin{equation}
d\Phi=\int
\frac{
d^{3}k
}{
(2\pi)^{3}2E_{\chi^{0}}
}
\frac{
d^{3}k'
}{
(2\pi)^{3}2E_{\chi^{+}}
}
\cdot (2\pi)^{4}\delta (p_{t}-p_{b}-k-k')\ \ ,
\end{equation}
are the phase space of $\chi^0$,$\chi^+$ as they are on shell.
$k$, $k'$ are the four--momenta of $\chi^0$ and $\chi^+$, respectively.
%Equivalently,
%we can get the expression for $ImA_2$ by the replacement 
%$\frac{1}{k^{2}-m^{2}+i\epsilon}
%\rightarrow -2\pi i\ \delta(k^{2}-m^{2})$ for the propagators of $\chi^0$
%and $\chi^+$.
%This procedure has the advantage to keep signs
%consistent in different graphs compared to directly using Eq.~\ref{cut}.
After summing up all spins of external particles we get, e.g., the interference
term of Fig.2a and the tree--level graph of the form
\begin{eqnarray}
\label{main}
&&\frac{1}{2}\sum_{\mbox{spin}}\Delta|M|^{2}(a)=-2\sum_i Im(a^*b^i)
\sum_{\mbox{spin}}Im(A_2^i A_1^*\nonumber)\\
&&=\frac{g^{6}}{q^{2}-m_W^{2}}
\frac{1}{(2\pi)^{2}}\int\frac{d^{3}k}{E}\delta((q-k)^{2}-m^2_{\chi})\frac{
\sum_i {{\mathcal X^{i}_a}F_{i}}
}{
{\mathcal P_{1}}\ {\mathcal P_{2}}
}
\end{eqnarray}
where we have introduced ${\mathcal X^i_a}$s to represent quantities
like $Im(a^*b^i)$ in Eq.~\ref{aa} arising from the SUSY couplings
 and the corresponding
$F_i$s to represent quantities like $Im(A_2A_1^*)$
which are Lorentz invariant
 functions of the four--momenta of 
$\chi^0$, $b$, $\bar{\tau}$ and $\nu_\tau$. 
${\mathcal P_1}$ and ${\mathcal P_2}$ are 
denominators of the two boson propagators in the loop graphs.
Note that the SM phase from $V_{tb}$ is cancelled in the interference of the
tree diagram and the one--loop diagrams.

We get the analytic expressions for $\frac{1}{2}\sum_{spin}\Delta|M|^2$ by
integrating the phase space of $\chi^0$ and $\chi^+$ in $\vec{q}=0$ system.
The analytic
results are given in appendix C. By expressing the formulae in
Lorentz invariant form, we translate the formulae to the top quark rest system.
In this system the final state three
body phase space integration is implemented numerically.

All the ${\cal X}^i$s for each graphs are proportional to $\xi_t^j$
(for detail expressions of ${\cal X}^i$s, see Appendix C), so, we can
write
\begin{equation}
\label{rati}
A_{CP}=\xi_{t}^1\cdot f_{CP},
\end{equation}
where
\begin{equation}
\xi_t^j = Im{Z_{t}^{1j}}^{\ast}Z_{t}^{2j} 
= \frac{(-1)^j}{2}\sin (2\theta_{t})\sin \phi_{t},
\end{equation}
 $\theta_t$ and 
$\phi_t$ are given in (A.13) and (A.9) of Appendix A.
In terms of parameters in the top squark mass matrix we get that
\begin{equation}
\xi_t^j =\frac{(-1)^{j-1} m_t\cdot ImA_t}{\sqrt{\Delta}}
\end{equation}
From the expression of $\Delta$ in (A.13) of Appendix A we can see that 
$|\xi_t^i|$ can be as large as $\frac{1}{2}$ when the following conditions are
satisfied at the same time, i.e., $L_f=R_f$, $\mu=0$ and $A_t$ is purely
imaginary. This is certainly difficult to reach.

\section {Numerical results}
We now turn to our main numerical results.
The calculation is based on the low energy MSSM scenario whose parameter freedom
has been greatly reduced as described in section 2. Another simplifying 
assumption taken in our calculation is the 
universal relationship between the gaugino masses, i.e.,
${m_1}=\frac{5}{3}{m_2}
\tan^2\theta_{W}$, where $\theta_{W}$ is the 
weak mixing angle\cite{bb}.
We write the parameters $m^2_{\tilde{t}L}$, $m^2_{\tilde{t}R}$ in the
top squark mass matrix as
\begin{eqnarray}
m^2_{\tilde{t}L}=M^2-cm^2_t,&m^2_{\tilde{t}R}=M^2-2cm^2_t
\end{eqnarray}
where ${M}$ is an arbitrary mass scale for scalar particles.

Neglecting the masses of $\tau$ and quarks except top quark we are left with
ten SUSY parameters, i.e., $\mu$, $m_2$,
$\tan\beta$, c, ${M}$, $|A_t|$, $\arg(A_t)$, $m_{\tilde{\tau}L}$
,$m_{\tilde{\tau}R}$ and $m_{\tilde{\nu}}$.
We take  
$m_{\tilde{\tau}L}=m_{\tilde{\tau}R}=m_{\tilde{\nu}}=130 GeV$ to which
the results are insensitive, and always take $|A_t|=M$,
$c=0.1\sim 1$.
The other free SUSY parameters are restricted by the experimental limits on
the masses
of super particles \cite{pdg} and our assumption $m_{\tilde{t_1}}\geq 140 GeV$.
 In particular, we take that,
$m_{{\chi^0}_1}$, the mass of the lightest neutralino, is above 30GeV and
$m_{{\chi^+}_1}$, the mass of the light chargino, is above 65GeV. Another 
limit for the parameter space is adopted for simplicity that we
require $m_{{\chi^0}_1} + m_{{\chi^+}_1} > 100GeV$ and 
$m_{{\chi^0}_1}+m_{\tilde{t}_1} > m_t$.
The SM parameters are taken as $m_t=175GeV$, $|V_{tb}|^2=1$,
$\alpha=\frac{1}{128}$, $m_W=80.33GeV$ , 
$\sin\theta_W=0.232$ and $m_b=m_\tau=0$.

A consequence of the above scenario, especially that
$\arg(\mu)$=0 and $m_b=0$ is that Figs.2c, 2d and 2f do not acquire 
any CP--violating phase. So only Figs.2a, 2b and 2e are considered. It
is found numerically that more than 90\% of contributions to $A_{CP}$ come
from the triangle diagram Fig.2e. Because of this, the results are not 
sensitive to the
values of $m_{\tilde{\tau}}$ and $m_{\tilde{\nu}}$.

We have studied
the CP asymmetry, $A_{CP}$, as a function of SUSY parameters $\arg(A_t)$,
$\mu$, $\tan\beta$, $m_2$ and $m_{\tilde{t}_1}$. In all the
figures there are two curves for the same values of the fixed parameters of 
which the curve giving
larger $A_{CP}$ represents
$A_{CP}^t$ while that giving a smaller value represents $A_{CP}^e$.

In Fig.3 we show the $A_{CP}$ as a function of $\arg(A_t)$. We can see that 
$A_{CP}$ is approximately a sine function of $\arg(A_t)$
as can be seen in Eq.~\ref{rati} and Eq. 19.
$f_{CP}$ defined in Eq.~\ref{rati}
is plotted in Fig.4. We see that $f_{CP}$ is just like a parabola.
As a result
$A_{CP}$ does not reach its maximum when $A_t$ is purely imaginary, rather, it
is maximal at $\arg(A_t)\approx \pm0.7\pi$. 
$f_{CP}$ depends on $\arg(A_t)$ through the top squark mass (see Appendix A).

In Figs.5--7, we plotted $A_{CP}$ as a function of the Higgs mass parameter 
$\mu$
for $\tan\beta=1.2, 5, 15$ respectively, for different values of $m_2$. 
The global feature of the three figures is that $A_{CP}$ decrease dramatically
as $|\mu|$ increase.
Notice that, in the high $\tan\beta$ scenario $A_{CP}$ becomes quite insensitive
to the sign of $\mu$. $A_{CP}$ is almost symmetric about $|\mu|$ for $\tan\beta=15$.
However,
as $\tan\beta=1.2$ $A_{CP}$ is not small only for minus $\mu$.
Another feature of the figures is
that for fixed $\tan\beta$ and $\mu$, $A_{CP}^e$ decrease whereas $A_{CP}^t$
increase as $m_2$ become larger. The reason is evidently that the 
threshold varies with $m_2$.

The dependence of $A_{CP}$ on $\tan\beta$ is plotted in Fig.8 and Fig.9,
for several values of $m_2$ and for 
$\mu=-70GeV$ and $\mu=-50GeV$, respectively.
An interesting feature of these two figures is that $A^t_{CP}$ decreases as
$\tan\beta$ increases whereas $A^e_{CP}$ increases. This is because low
$\tan\beta$ gives strong Yukawa coupling for top quark 
so that we get large $A^t_{CP}$.
However, large $\tan\beta$ makes the threshold lower and thus elevates 
$A^e_{CP}$. So,
$A^e_{CP}$ prefers bigger $\tan\beta$ in contrast to the situation in the top
quark two
body decays.
As $\tan\beta > 4$, we can see from Fig.8 that $A_{CP}^t$ is almost
insensitive to $\tan\beta$.

The dependence of $A_{CP}$ on $m_2$ is plotted in Fig.10, for several values
of $\tan\beta$.
We can see that the major part of the curve $A_{CP}^e$ falls into the 
region between 0.1\%--0.2\%. 
For $\tan\beta=2$, $A_{CP}$ rises as $m_2$ increases whereas
for large $\tan\beta$ $A_{CP}$ slightly drops as $m_2$ increases.

Finally, we give the dependence of $A_{CP}$ on the top squark mass in Fig.11
for $\mu=-50GeV$. It is found
that $A_{CP}$ depends essentially only on $m_{\tilde{t}_1}$, not separately
on ${M}$ and c.  $A_{CP}$
decreases with $m_{\tilde{t}}$, just as what has been expected. When
$m_{\tilde{t}_1}$ is around 140 GeV $A_{CP}$ can reach up to 0.5\%.

In summary, $A^e_{CP}$ is around 0.1\% level in the major part of the
 parameter space that we have 
discussed. In a very narrow region of the parameter space, 
$A^e_{CP}$ can reach 0.5\% level.

\section {Discussion}
We would like to point out two points in this section.

(1). The branch ratio for top quark decay drops rapidly when W boson
is off shell.
For $\sqrt{q^2} > 100GeV$ the branch ratio for $b\nu_\tau\tau$
final state is only about $\frac{1}{1000}$
according to our calculation. However, as we have mentioned that the
 contributions to
$A_{CP}$  comes mainly from the triangle diagrams in Fig.2, $A_{CP}$ for
different final states have the same sign and approximately the same size. 
So we can make a combining analysis for the data of
the three body CP asymmetries for different three fermion final states
($\tau,\nu_\tau$), ($\mu,\nu_\mu$), ($e,\nu_e$), and all allowed three quark
 decays . The total branch ratio 
for top quark three body decay for $\sqrt{q^2} > 100GeV$ can reach up to
about $\frac{1}{100}$. However, there are CP asymmetries of the order
$(\frac{\alpha_s}{\pi})^2$ for three quark decays coming from gluon and
gluino corrections, which may be of the same order as considered here. This
needs to be studied further if one attempts to quantitatively compare the theory
and experimental CP asymmetries for three quark decays.

(2). The total width of top and anti--top quark are 
equal $\Gamma=\bar{\Gamma}$ due to CPT theorem. The following relation holds
\begin{equation}
\label{eq}
\Gamma(t\rightarrow bf\bar{f'})-\bar{\Gamma}
(\bar{t}\rightarrow\bar{b}\bar{f}f')=-\left(\ 
\Gamma(t\rightarrow b\chi^{0}\chi^{+})-\bar{\Gamma}
(\bar{t}\rightarrow\bar{b}\chi^{0}\chi^{-})\ \right)_{f\bar{f'}},
\end{equation}
where $f\bar{f'}$ represent all the W boson decay products mentioned above 
and the RHS of Eq.~\ref{eq} denotes the CP asymmetry from the diagrams with 
$f\bar{f'}$ as intermediate states.
The relation can be easily seen from the Feynman diagrams representing these
two processes in Fig.1, 2 and Fig.12.

The four interference terms contributing to the RHS of Eq.~\ref{eq} 
between the two triangle diagrams and the two tree--level diagrams 
in Fig.12 correspond to the interference terms between the four box diagrams
in Fig.2 and the tree--level diagram in Fig.1. The other two
interference terms of the fermion loop diagram with the two tree--level
graphs in Fig.12
correspond to those of the triangle
diagrams in Fig.2 with the tree--level diagrams in Fig.1. The relative minus 
sign on the RHS of
Eq.~\ref{eq} can be explained as following.
We must add a minus sign to the amplitude
for Fig.12e
due to the closed fermion loop. Such a minus sign is absent
in the corresponding amplitude in Fig.2 .
On the contrary, there is a minus sign in front of the
box diagram in Fig.2. No such minus sign is present in Fig12.c, d.
Thus, the CPT relation Eq.~\ref{eq} manifests itself through the
Cutkosky rule in
Eq.~\ref{cut}.

\section {Summary}
In this work we considered the CP asymmetries induced by MSSM new phase in
the semi--leptonic three body decays of the top quark under such an assumption 
that the top squark is
so heavy that no CP--odd effects are observable in top quark two body decays
in one--loop level. In our calculation $A_{CP}$ can reach up to 0.55\% in the
most favorable case. Considering the small total branch ratio for
$\sqrt{q^2}>100GeV$ which is about $\frac{1}{100}$ for all three body decays
it is really hard to detect such small effects experimentally.

However, several constraint in our calculations can be relaxed. For example,
the gaugino mass parameters $m_1$ and $m_2$ can be complex
and give new CP--violating sources if we do not make the universal assumption
about the gauginos masses in GUT scale. Another possible CP source is from
the Higgs mass parameter $\mu$. According to recent studies on the neutron EDM,
cancellation among different contributions can take place so that $\mu$
can have large imaginary part even when super particles are in
${\cal O}(100GeV)$ level\cite{hp}. If this is the case, complex $\mu$ can
introduce additional CP asymmetries in top quark decays. And, if we relax the
constraint on the $m_{\tilde{t}_1}$ to be slightly above 100GeV, 
$m_{\tilde{t}_1}$ 
and $m_{\chi^0}$ cut can also give contribution to $A_{CP}$ in three body
decays. Another improvement may come from branch ratio enhancement. The direct
experimental constraint on $m_{\chi^0}+m_{\chi^+}$ is even below $m_W$ today
\cite{pdg}. As $\sqrt{q^2}$ decrease the branch ratio for top quark three 
body decay increase rapidly (e.g., the branch ratio for all three body decays
is about $\frac{1}{35}$
for $\sqrt{q^2} > 90GeV$). So, if $m_{\chi^0}+m_{\chi^+}$ is not much heavier
than $m_W$, the $A_{CP}$ in the three body decay is hopefully detectable in
the LHC which may be able to produce $10^7$ $t\bar{t}$ pairs.

On the contrary, if $m_{\chi^+}$ and $m_{\tilde{t}}$ are both
heavier than, e.g., 140GeV, all the windows for CP asymmetries  
induced by
super particles will be shut up. CP--odd effects in top quark decays 
induced by MSSM particles can only exist
beyond one--loop order. This will be beyond the experiment ability in the
near future.

\begin{center}
{\bf Appendix A}\\
\end{center}

There are different conventions for super particles mass matrices 
adopted in literature so that it is very easy to make
sign errors. To avoid sign errors we rederived the MSSM Lagrangian. For most
part we 
adopted the same conventions described by J. Rosiek\cite{j}. The original 
MSSM Lagrangian is given
by Eq.~\ref{1}---Eq.~\ref{10} of which all fields are gauge eigenstates.
We should point out that the $\epsilon_{12}=-1$
convention will give an extra minus sign to parameter $\mu$ compared to those
adopted  the convention $\epsilon_{12}=1$(if universal relation
for gauginos is taken only relative sign between the gauginos and
$\mu$ is significant). To get the physical spectrum of
particles one should carry out the standard procedure of gauge symmetry 
breaking. After SSB super--particles will mix and form different mass 
eigenstates.

\setcounter{num}{1}
\setcounter{equation}{0}
\def\theequation{\Alph{num}.\arabic{equation}}

The charged Higgsinos and charged winos mix and give two mass eigenstates named
charginos.
The mass matrix of charginos is
\begin{equation}
M_{\chi} = \left[\matrix{
{m}_2 & \sqrt{2} M_W \sin\beta \cr
\sqrt{2} M_W \cos\beta & \mu } \right] \ \ .
\end{equation}
The mixing matrices satisfy
\begin{equation}
(Z^-)^TM_{\chi} Z^+
= {\rm diag}\left(m_{{\chi}_1}, m_{{\chi}_2}\right) \ ,
\end{equation}
and is defined by
\begin{equation}
\left( \begin{array}{c}
 -i\ \lambda^{-}\\ \psi^{2}_{H_1} 
\end{array} \right)\ = \ 
Z^{-}\ \left(\begin{array}{c}\varphi^{-}_{1}\\ \varphi^{-}_{2}\end{array}
\right)\ \ ,
\end{equation}

\begin{equation}
\left( \begin{array}{c}
 -i\ \lambda^{+}\\ \psi^{1}_{H_2} 
\end{array} \right)\ = \ 
Z^{+}\ \left(\begin{array}{c}\varphi^{+}_{1}\\ \varphi^{+}_{2}\end{array}
\right)\ \ .
\end{equation}
In the above equations 
$\lambda^{\pm}=\frac{1}{\sqrt{2}}(\lambda^{1}_{W}\mp i\lambda^{2}_{W})$ where
$\lambda^{1,2}_{W}$ are the first and second components of wino as in 
Eq.~\ref{gau}.
$\psi^{2}_{H_1}$ is the first (or up) Fermion component of the second Higgs 
super field doublet.
The fields on the left hand side of the above equations are gauge eigenstates and
the fields on the right hand side are mass eigenstates.
The four--component Dirac spinor charginos are defined by $\chi_{i}^{+}=\left[
\begin{array}{c}\varphi^{+}_{i}\\ \bar{\varphi^{-}_{i}}\end{array}\right]$.
The mass term
which will appear in the final form of Lagrangian is $-m_{\chi_i}\bar{\chi_i}
\chi_i$. 

The third component of wino, photino and neutral Higgsinos combine to give
four Majarana neutralinos.
The mass matrix for neutralinos is
\begin{equation}
M_{{\chi}^0} = \left[\matrix{
{m}_1   & 0 & -M_Z\cos\beta\sin\theta_W & M_Z\sin\beta\sin\theta_W \cr
0 & {m}_2 & M_Z\cos\beta\cos\theta_W & -M_Z\sin\beta\cos\theta_W \cr
-M_Z\cos\beta\sin\theta_W & M_Z\cos\beta\cos\theta_W & 0 & -\mu\cr
M_Z\sin\beta\sin\theta_W & -M_Z\sin\beta\cos\theta_W & -\mu & 0 } \right] \ , \ \ 
\end{equation}
which is diagonalized by

\begin{equation}
Z_N^T M_{{\chi}^0} Z_N
= {\rm diag}\left(m_{{\chi}^0_1}, m_{{\chi}^0_2},
m_{{\chi}^0_3}, m_{{\chi}_4^0}\right) \ \ .
\end{equation}
$Z_N$ is defined by

\begin{equation}
\left( \begin{array}{c}
-i\lambda_{B}\\ -i\lambda_{W}^3\\ \psi^{1}_{H_1}\\ \psi^{2}_{H_2} \end{array}
\right) \ =\ Z_{N}\left( \begin{array}{c}
\varphi^{0}_{1}\\ \varphi^{0}_{2}\\ \varphi^{0}_{3}\\ \varphi^{0}_{4} 
\end{array}\right) \ \ ,
\end{equation}
where $\lambda_B$ is photino. All fields on the left hand side of the above
equation are gauge eigenstate and those on the right hand side are mass 
eigenstates.
The four--component Dirac spinor form for neutralino is $\chi^{0}_{i}=
\left[\begin{array}{c}
\varphi^{0}_{i}\\ \bar{\varphi^{0}_{i}}\end{array}\right]$.
The mass term in the final form of Lagrangian of neutralino is $-\frac{1}{2}
m_{\chi^0_i}\bar{\chi^0_i}\chi^0_i$.

Ignoring generation mixing, the mass eigenstates of squarks are obtained by
mixing the left--handed and right--handed eigenstates of squarks.
Its mass matrix is 
\begin{eqnarray}
M^2_{\tilde{f}}& =& \left[\matrix{
L_f
&C_f  \cr
C_f^*
&R_f }\right]\\
%\end{equation}
%\begin{eqnarray}
L_f&=&m^2_f + \cos 2 \beta(T_{3f} - Q_f \sin^2 \theta_W)M_Z^2 + 
m^2_{\tilde{f}_L}\nonumber\\
R_f&=&m^2_f+\cos 2\beta Q_f \sin^2 \theta_WM_Z^2+m^2_{\tilde{f}_R}\nonumber\\
C_f&=&- m_f(r_f\mu + A_f^*)=|C_f|e^{i\phi_f}
\end{eqnarray}
where $T_{3f}$ is $\frac{1}{2}$ for up squark and $-\frac{1}{2}$ 
for down squark.
$Q_f$ is the charge of the sparticle and $r_f$ is $\cot\beta$ for up squark and
$\tan\beta$ for down squark. $m^2_{\tilde{f}_L}$, $m^2_{\tilde{f}_R}$, $A_f$
are the corresponding diagonal elements of $m^2_Q$, $m^2_U$($m^2_D$) and
$A_U$($A_D$) in ${\cal L}_{\rm soft}$, respectively.
The mixing matrix satisfies
\begin{equation}
Z^{\dagger}_f M^2_{\tilde{f}} Z_f
= {\rm diag}\left(m^2_{\tilde{f}_1}, m^2_{\tilde{f}_2}\right).
\end{equation}
They are defined, e.g., for up squark, by
\begin{equation}
\label{dd}
\left(
\begin{array}{c}
\tilde{Q}^U_L \\ \tilde{U}^*_R
\end{array}
\right) =
Z_{U}\left(
\begin{array}{c}
\tilde{U}_1 \\ \tilde{U}_2
\end{array}
\right),
\end{equation}
where $\tilde{Q}^U_L$ is the up component of left--handed up squark doublet,
 $\tilde{U}_R$ is 
the right--handed up squark. $\tilde{U}_{1,2}$ are the two mass eigenstates of
up squark. The conjugate of $\tilde{U}_R$ in Eq.~\ref{dd}
comes from that we adopt the charge
conjugate of left hand Fermion to represent its right hand component
in the original MSSM Lagrangian. The final
form of squark mass term in Lagrangian is 
$-m^{2}_{U_i}\tilde{U_i}^{*}\tilde{U_i}$.
In particular, the mixing matrix for top squark is given by
\begin{eqnarray}
Z_{t}& =& \left[\matrix{
\cos\theta_te^{i\phi_t/2}
&-\sin\theta_te^{i\phi_t/2} \cr
\sin\theta_te^{-i\phi_t/2} 
&\cos\theta_te^{-i\phi_t/2} }\right]\ \ ,
\end{eqnarray}
where $\phi_t$ is defined in (A.9) and 
\begin{eqnarray}
\tan\theta_t&=&\frac{2|C_t|}{L_t-R_t-\sqrt{\Delta}}\ \ , \nonumber\\
\Delta&=&(L_t-R_t)^2+4|C_t|^2
\ \ .
\end{eqnarray}
The formulae for down squark are similar to up squark. Note the definition of
mixing matrix for down squark given by \cite{j} is the complex conjugate of 
the mixing matrix given here.

\begin{center}
{\bf Appendix B}\\
\end{center}
\setcounter{num}{2}
\setcounter{equation}{0}
\def\theequation{\Alph{num}.\arabic{equation}}

In this appendix we list the relevant pieces of the SUSY Lagrangian 
in terms of the mass eigenstates\cite{j}. 

\begin{eqnarray}
{\cal L}_{t\tilde{t}\chi^0}&=& g\tilde{t}^*_{i}\bar{\chi^0_{j}}[A^{ij}P_{L}
+B^{ij}P_{R}]t\ +\ H.C.\\
{\cal L}_{b\tilde{t}\chi}&=&g\tilde{t}_{i}\bar{b}[C^{ij}P_{L}+D^{ij}P^{R}]
V_{tb}^{*}\chi^{c}_{j}+H.C.\\
{\cal L}_{b\tilde{b}\chi^0}&=&g\tilde{b}_i^{*}\bar{\chi^0_j}[E^{ij}P_{L}+F^{IJ}
P_{R}]b+H.C.\\
{\cal L}_{t\tilde{b}\chi}&=&g\tilde{b}_i^{*}\bar{\chi_j}[G^{ij}P_{L}+H^{ij}P_{R}]
V_{tb}^{*}t+H.C.\\
{\cal L}_{\tau\tilde{\tau}\chi^0}&=&g\tilde{\tau}^{*}_{i}\bar{\chi^0_j}
[M^{ij}P_{L}+N^{ij}P_{R}]\tau+H.C.\\
{\cal L}_{\nu\tilde{\tau}\chi}&=&gU^{ij}\tilde{\tau}_{i}^{*}\bar{\chi_j}P_{L}
\nu_{\tau}+H.C.\\
{\cal L}_{\tau\tilde{\nu}\chi}&=&-g\tilde{\nu}_{\tau}^{*}\bar{\chi^c_i}
[Z^+_{1i}P_{L}+l^{\tau}{Z^{-}_{2i}}^{*}P_R]+H.C.\\
{\cal L}_{\nu\tilde{\nu}\chi^0}&=&gW_{i}\tilde{\nu}_{\tau}^{*}
\bar{\chi^0_i}P_{L}\nu_{\tau}+H.C.\\
\label{kkw}
{\cal L}_{\chi\chi^{0}W}&=&g\bar{\chi_i}\nu^{\mu}[O^{ij}P_{L}+V^{ij}P_{R}]
\chi^{0}W^+_{\mu}+H.C.
\end{eqnarray}

where
\begin{equation}
\label{b10}
\begin{array}{l}

\left\{ 
\begin{array}{l}
A^{ij}=\frac{-1}{\sqrt{2}\cos\theta}{Z^{1i}}^{*}_{t}(\frac{1}{3}Z^{1j}_{N}
\sin\theta+Z^{2j}_{N}\cos\theta)-u^{t}{Z_t^{2i}}^{*}Z_N^{4j}\\
B^{ij}=\frac{2\sqrt{2}\sin\theta}{3\cos\theta}{Z_{t}^{2i}}^{*}{Z_N^{1j}}^*-
u^{t}{Z_t^{1i}}^{\ast}
{Z_N^{4j}}^{\ast}
\end{array} \right. \\
\left\{
\begin{array}{l}
C^{ij}=-d^{b}Z_{t}^{1i}Z_{2j}^{-}\\
D^{ij}=-Z_{t}^{1i}{Z_{1j}^{+}}^{*}+u^{t}Z_{t}^{2i}{Z_{2j}^{+}}^*
\end{array}\right. \\
\left\{
\begin{array}{l}
E^{ij}=\frac{-g}{\sqrt{2}\cos\theta}Z_b^{1i}(\frac{1}{3}Z_N^{1j}\sin\theta
-Z_N^{2j}\cos\theta)+d^{b}Z_b^{2i}Z_N^{3j}\\
F^{ij}=\frac{-\sqrt{2}\sin\theta}{3\cos\theta}Z_b^{2i}{Z_N^{1j}}^{*}+d^{b}
Z_b^{1i}{Z_N^{3j}}^{*}\end{array}\right. \\
\left\{
\begin{array}{l}
G^{ij}=-Z_b^{1i}Z^{-}_{1j}-d^{b}Z_b^{2i}Z^{-}_{2j}\\
H^{ij}=u^{t}Z_b^{1i}{Z^{+}_{2j}}^{*}
\end{array}\right. \\
\left\{
\begin{array}{l}
M^{ij}=\frac{1}{\sqrt{2}\cos\theta}Z_{\tau}^{1i}(Z_N^{1j}\sin\theta+Z_N^{2j}
\cos\theta)+l^{\tau}Z_{\tau}^{1i}Z_N^{3j}\\
N^{ij}=\frac{-\sqrt{2}\sin\theta}{\cos\theta}Z_{\tau}^{2i}{Z_N^{1j}}^{*}
+l^{\tau}Z_{\tau}^{1i}{Z_N^{3j}}^{*}
\end{array}\right. \\
U^{ij}=-(Z_{\tau}^{1i}Z^{-}_{1j}+l^{\tau}Z_{\tau}^{2i}Z^-_{2j}) \\
W_{i}=\frac{1}{\sqrt{2}\cos\theta}(Z_N^{1i}\sin\theta-Z_N^{2i}\cos\theta) \\
\left\{
\begin{array}{l}
O^{ij}=Z_N^{2i}{Z^+_{1j}}^{*}-\frac{1}{\sqrt{2}}Z_N^{4j}{Z^+_{2j}}^{*}\\
V^{ij}={Z_N^{2i}}^{*}Z^-_{1j}+\frac{1}{\sqrt{2}}{Z_N^{3i}}^{*}Z^-_{2j}
\end{array}\right.
\end{array}
\end{equation}
$u^t$, $d^b$ and $l^\tau$ are the Yukawa coupling parameters for top,
bottom and $\tau$ respectively. They are given by
\begin{eqnarray}
u^t&=&\frac{m_t}{\sqrt{2}m_W\sin\beta}\ \ \ ,\nonumber\\
d^b&=&\frac{-m_b}{\sqrt{2}m_W\cos\beta}\ \ \ ,\nonumber\\
l^\tau&=&\frac{-m_\tau}{\sqrt{2}m_W\cos\beta}\ \ \ .
\end{eqnarray}

The vertex in Eq.~\ref{kkw} can be expressed as
\begin{equation}
{\cal L}_{\chi\chi^{0}W}=-g\bar{\chi^c_{j}}\nu^{\mu}
[{O^{ij}}^{*}P_{R}+{V^{ij}}^{*}P_{L}]\chi^0_{i}W_{\mu}^{-}+H.C.
\end{equation}
which can be used to read off the Feynman rule directly for the Fig.2e.

\begin{center}
{\bf Appendix C}\\
\end{center}
\setcounter{num}{3}
\setcounter{equation}{0}
\def\theequation{\Alph{num}.\arabic{equation}}
\def\pt {p_{t}}
\def\pb {p_{b}}
\def\pg {p_{\nu}}
\def\ptau {p_{\tau}}
\def\cf {{\cal X}_a^1}
\def\c2 {{\cal X}_a^2}
\def\ef {{\cal X}_e^1}
\def\es {{\cal X}_e^2}
\def\et {{\cal X}_e^3}
\def\eff {{\cal X}_e^4}
\def\en {E_{\nu_\tau}}

In this section we give the formulae for the quantity $\frac{1}{2}
\sum\Delta |M|^2$ in Eq.~\ref{main} for Fig. 2c and 2e in detail.
In the following formulae the ${\cal X}^i$s are not separated out explicitly
as in Eq.~\ref{main} for convenience. All the formulae and variables
are given in the $\vec{q}=0$ system.
The imaginary part of the amplitude for Fig2.c is
\begin{eqnarray}
&&ImA(c)=\frac{g^4}{2(2\pi)^2}\int\frac{d^3k}{2E}\delta((q-k)^{2}-m^2_{\chi})
\cdot\nonumber\\
&&\frac{
\bar{u}(p_{\nu})\Gamma_{\tilde{\nu}\nu\chi^0}
(k\hspace{-0.2cm}/+m_{\chi^0})\Gamma_{\tilde{t}t\chi^0}u(p_t)
\bar{u}(p_{b})\Gamma_{\tilde{t}b\chi}(q\hspace{-0.2cm}/-k\hspace{-0.2cm}/
-m_{\chi})\Gamma_{\tilde{\nu}\tau\chi}
v(p_{\tau})
}
{
[(\pt-k)^{2}-m^2_{\tilde{t}}]\ [{(p_{\nu}-k)}^{2}-m^2_{\tilde{\nu}}]
}
\end{eqnarray}
where $k=(E,\vec{k})$ is the four-momenta of $\chi^0$ and the $\Gamma$s
are interaction vertex factors. 
The quantity $\frac{1}{2}\sum_{spin}\Delta|M|^2$ for Fig.2c is
\begin{eqnarray}
\frac{1}{2}\sum\Delta|M|^2(c)&=&
\frac{g^6}{q^2-m^2_W}\frac{|\vec{k}|}{4\pi\sqrt{q^2}}
\left\{
\frac{1}{A^{'}a}(2T+L\en^{'}|\vec{k}|\cos\alpha)+
\frac{2\rho b\cos\alpha} {A^{'}B^{'}a}\nonumber \right.\\
&& -\frac{a(S+LB^{'}/4)+T+\frac{1}{2}L\en^{'}|\vec{k}|\cos\alpha}
{A^{'}a^2}\log\frac{1+a}{1-a}\nonumber \\
&& +\frac{1}{A^{'}B^{'}ab}(-\Omega+\rho(\frac{1}{a}+\frac{1}{b}\cos\alpha))\log
\frac{1-b}{1+b}\nonumber\\
&&\left.
+\frac{1}{A^{'}B^{'}}(\Sigma-\frac{\Omega}{a}+\frac{\rho}{a^2})\frac{1}{\sqrt{K}}
\log\frac{1-ab\cos\alpha+\sqrt{K}}{1-ab\cos\alpha-\sqrt{K}} \right\}
\end{eqnarray}
where
\begin{eqnarray}
\rho&=&-2\c2 \pg \cdot \ptau |\vec{k}|^{2}|\vec{p_t}|^{2}\nonumber\\
\Omega&=&(\cf m_{\chi^0}m_{t}-2\c2 (\pb  \cdot \pg +k\cdot
q))\pg \cdot \ptau |\vec{k}||\vec{p_t}|\nonumber\\
&&+4\c2 \pg \cdot \ptau EE_{t}|\vec{k}||\vec{p_t}|
-2T^{'}\en^{'}E-B^{'}T\nonumber\\
\Sigma&=&2\cf m_{\chi^0}m_{t}\pb  \cdot \pg \pg \cdot \ptau 
+(2\c2 (\pb  \cdot \pg +k\cdot q)-\cf m_{\chi^0}m_{t})
\pg \cdot \ptau EE_{t}\nonumber\\
&&+\cf m_{\chi^0}m_{t}\pg \cdot \ptau k\cdot q-2\c2 
\pg \cdot \ptau (EE_{t})^{2}\nonumber\\
&&+2S^{'}\en^{'}E-L\en^{'}E^{2}+B^{'}S+LB^{'}/4\nonumber\\
&&-\c2 (\pb  \cdot \pg \pt\cdot \ptau -\pb  \cdot \pt\pg \cdot 
\ptau +\pt\cdot \pg \pb  \cdot \ptau )\nonumber\\
K&=&a^2+b^2-a^2b^2\sin^2\alpha-2ab\cos\alpha
\end{eqnarray}
In above formulae, $E_{\nu_\tau}^{'}$ is the energy of 
$\pg$ in the $\vec{q}=0$ system.
$\alpha$ is the angle between the three--momenta of top quark and $\nu_\tau$
in the $\vec{q}=0$ system. $q^2$ is the invariant mass square of the final state
lepton pair.
Other quantities are defined as:
\begin{eqnarray}
A&=&m_t^{2}+m^2_{\chi^0}-m^2_{\tilde{t}},\nonumber\\
B&=&m^2_{\chi^0}-m^2_{\tilde{\nu}},\nonumber\\
A^{'}&=&A-2EE_{t},\nonumber\\
B^{'}&=&B-2E\en^{'},\nonumber\\
a&=&\frac{2|\vec{k}||\vec{p_t}|}{A^{'}},\nonumber\\
b&=&\frac{2\en^{'}|\vec{k}|}{B^{'}},\nonumber\\
L&=&\c2 \pt\cdot \pb  ,\nonumber\\
S&=&\frac{1}{2}\cf m_{\chi^0}m_{t}\pb  \cdot q+\c2 \pb  \cdot
\pg \pt\cdot q+\nonumber\\ &&\c2 (\pb  \cdot \ptau -\pt\cdot \pg )EE_{t}+
\c2 \pt\cdot \pg k\cdot q,\nonumber\\
T&=&\c2 (\pb  \cdot \ptau -\pt\cdot \pg )|\vec{k}||\vec{p_t}|
,\nonumber\\
S^{'}&=&\frac{1}{2}\cf m_{\chi^0}m_{t}(\pb  \cdot \ptau -\pb  \cdot \pg) 
+\c2 \pb  \cdot\pg \pt\cdot q+\nonumber\\
&&\c2 (\pb  \cdot \ptau +\pt\cdot \pg )EE_{t}-
\c2 \pt\cdot \pg k\cdot q,\nonumber\\
T^{'}&=&\c2 (\pb  \cdot \ptau +\pt\cdot \pg )|\vec{k}||\vec{p_t}|
\end{eqnarray}
The non--Lorentz invariant four--vector component are expressed as
\begin{eqnarray}
\en{'}&=&\frac{\sqrt{q^2}}{2},\nonumber\\
E_t&=&\frac{m^2_{t}+q^2}{2\sqrt{q^2}},\nonumber\\
|\vec{p_t}|&=&\frac{m^2_{t}-q^2}{2\sqrt{q^2}},\nonumber\\
E&=&\frac{m^2_{t}+q^2}{2m_t},\nonumber\\
|\vec{k}|&=&\sqrt{E^2-m^2_{\chi^0}},\nonumber\\
\cos\alpha&=&\frac{m^2_{t}+q^2-4\pt\cdot \pg }{m^2_t-q^2}
\end{eqnarray}
${\cal X}_a^{1,2}$ are given by
\begin{equation}
{\cal X}_a^1=-ImB^{i1}D^{i1}Z^+_{11}{W^1}^*\ ,\ \
{\cal X}_a^2=-ImA^{i1}D^{i1}Z^+_{11}{W^1}^*
\end{equation}
The elements of the mixing matrices in the above equation are taken
according to that only $m_{\chi^0_1}$ and $m_{\chi^+_1}$ are 
considered in our calculations.

The corresponding formulae for Fig2.e are
\begin{eqnarray}
&&ImA(e)=\frac{g^4}{2\sqrt{2}(2\pi)^2}\int\frac{d^3k}{2E}
\delta((q-k)^{2}-m^2_{\chi})\nonumber\\
&&\cdot\frac{
\bar{u}(p_{b})\Gamma_{\tilde{\tau}b\chi}(k\hspace{-0.2cm}/-q\hspace{-0.2cm}/
+m_{\chi})\gamma^{\mu}
\Gamma_{\chi\chi^0W}\Gamma_{\tilde{t}t\chi^0}u(p_t)
\bar{u}(p_{\nu})\gamma_{\mu}P_{L}v(p_{\tau})
}
{
(q^2-m_W^2)[(\pt-k)^{2}-m^2_{\tilde{t}}]
}
\end{eqnarray}
The quantity $\frac{1}{2}\sum_{spin}\Delta|M|^2$ for fig.2e is
\begin{eqnarray}
\frac{1}{2}\sum\Delta|M|^2&=&\frac{-g^6}{(q^2-m_W^2)^2}\frac{1}{2\sqrt{2}\pi}
\frac{|\vec{k}|}{\sqrt{q^2}}\cdot\nonumber\\
&&\frac{1}{A^{'}a}\left\{
2(Y-\frac{X}{a})+(Z-\frac{Y}{a}+\frac{X}{a^2})\log\frac{1+a}{1-a}\right\}
\end{eqnarray}
where
\begin{eqnarray}
X&=&2\et [\en^{'}|\vec{P_t}|\cos\alpha(\pb  \cdot \ptau -\pt\cdot \pg )
-|\vec{P_t}|^{2}\pg \cdot \ptau +\nonumber\\
&&\frac{1}{2}{\en^{'}}^{2}\cos^{2}\alpha
\pt\cdot \pb  ]|\vec{k}|^2\nonumber\\
Y&=&\Omega-S|\vec{P_t}||\vec{k}|+T\en^{'}\cos\alpha\nonumber\\
Z&=&\Sigma+SEE_t+T^{'}E\en^{'}+H\nonumber\\
S&=&(2\et \pb  \cdot \pg -\eff m_{\chi^0}m_{t})
\pg \cdot \ptau \nonumber\\
T&=&2\es m_{\chi}m_{t}\pb \cdot \pg -\eff m_{\chi^0}m_{t}
\pb  \cdot \pt-2\et q\cdot \pt \pb  \cdot \pg \nonumber\\
T^{'}&=&2\es  m_{\chi}m_{t}\pb  \cdot \pg -
\eff m_{\chi^0}m_{t}(\pb  \cdot \pg -\pb  \cdot \ptau )-\nonumber\\
&&2\et \pb  \cdot \pg (\pt\cdot \pg -\pt\cdot \ptau )\nonumber\\
\Omega&=&2\et \{
-\pb  \cdot \ptau E\en^{'}|\vec{k}|(|\vec{P_t}|+E_{t}\cos\alpha)+
\pt\cdot \pg EE^{'}_{\tau}|\vec{k}|(E_{t}\cos\alpha-|\vec{P_t}|)\nonumber\\
&&+(2EE_{t}-k\cdot q)|\vec{P_t}||\vec{k}|\pg \cdot \ptau -
k\cdot q \en^{'}|\vec{k}|\cos\alpha_{0}\cdot \pg \}\nonumber\\
\Sigma&=&2\et \{
E^{2}E_{t}\en^{'}(\pb  \cdot \ptau +\pt\cdot \pg )-
EE^{'}_{\tau}k\cdot q \pt\cdot \pg -\nonumber\\
&&(EE_t-k\cdot q)EE_{t}\pg \cdot \ptau -
\en^{'}(E^{2}-\frac{1}{2}|\vec{k}|^{2})\pt\cdot \pb  \}\nonumber\\
H&=&\eff m_{\chi^0}m_{t}\pg \cdot \ptau  k\cdot q +2(\ef
m_{\chi^0}m_{\chi}\pt\cdot \ptau +\eff m_{\chi^0}m_{t}\pg \cdot \ptau )
\pb  \cdot \pg \nonumber\\
&&-\et m^{2}_{\chi^0}(\pb  \cdot \pg \pt\cdot \ptau -\pb  \cdot 
\pt\pg \cdot \ptau +\pt\cdot \pg \pb  \cdot \ptau )
\end{eqnarray}
The parameters ${\cal X}_e^i$s are given by
\begin{equation}
\begin{array}{cc}
\ef=ImA^{i1}D^{i1}(-{V^{11}}^*),& 
\es=ImB^{i1}D^{i1}(-{V^{11}}^*),\\
\et =ImA^{i1}D^{i1}(-{O^{11}}^*),& 
\eff =ImB^{i1}D^{i1}(-{O^{11}}^*). 
\end{array}
\end{equation}
From the above expressions and the expressions for A, D in Eq.~\ref{b10}
we can see that the ${\cal X}$s are proportional
to $\xi_{t}^i=Im({Z_t^{1i}}^*Z_t^{2i})$ as pointed out in section 3.

\pagebreak

\newpage

\section*{Figure Captions}

\newcounter{FIG}
\begin{list}{{\bf FIG. \arabic{FIG}}}{\usecounter{FIG}}
\item
The tree--level Feynman diagram for the process $t\rightarrow b\nu_{\tau}
\bar{\tau}$.
\item
The SUSY induced CP--violating one--loop diagrams for the process
$t\rightarrow b\nu_{\tau}\bar{\tau}$.
\item
The CP asymmetry $A_{CP}$ is plotted as a function of $\arg(A_t)$ for
$\tan\beta=1.2$, $m_2=150GeV$, $\mu=-40GeV$, ${M}=200GeV$, c=0.2. 
When $m_{\tilde{t}_1}=150GeV$, $A_{CP}$ reaches its maximum.
\item
The quantity $f_{CP}$ defined in Eq.~\ref{rati} is plotted as
function of $\arg(A_t)$. All the parameters are the same as that of 
Fig.3.
\item
The CP asymmetry $A_{CP}$ is plotted as a function of SUSY parameter $\mu$,
for several values of $m_2$, for $\tan\beta=1.2$, ${M}=160GeV$, c=0.15, 
$\arg(A_t)=0.5\pi$.
\item
The CP asymmetry $A_{CP}$ is plotted as a function of SUSY parameter $\mu$,
for several values of $m_2$, $\tan\beta=5$. All the other parameters are
the same as that of Fig.3.
\item
The CP asymmetry $A_{CP}$ is plotted as a function of SUSY parameter $\mu$,
for several values of $m_2$, $\tan\beta=15$. All the other parameters are
the same as that of Fig.3.
\item
The CP asymmetry $A_{CP}$ is plotted as a function of $\tan\beta$, for several
values of $m_2$, $\mu=-70GeV$, ${M}=160GeV$, c=0.15, $\arg(A_t)=0.5\pi$.
\item
The CP asymmetry $A_{CP}$ is plotted as a function of $\tan\beta$, for several
values of $m_2$, $\mu=-50GeV$, ${M}=160GeV$, c=0.15, $\arg(A_t)=0.5\pi$.
\item
The CP asymmetry $A_{CP}$ is plotted as a function of ${m_2}$ for
${M}=160GeV$, c=0.15, $\arg(A_t)=0.5\pi$, $\mu=-50GeV$ for $\tan\beta=2$
and $\mu=-60Gev$ for $\tan\beta=5,10$.
\item
The CP asymmetry $A_{CP}$ is plotted as a function of the mass of the 
light top squark, for different values of $m_2$, $\tan\beta=2.5$,
$\mu=-50GeV$, $\arg(A_t)=0.5\pi$.
\item
The tree--level and one--loop Feynman diagrams for the process 
$t\rightarrow b\chi^0\chi^+$.
\end{list}
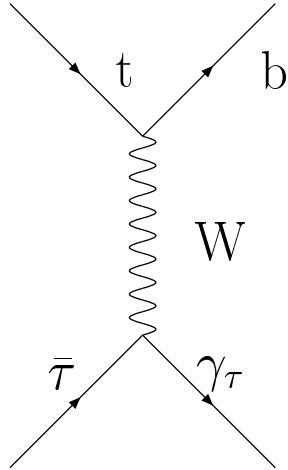
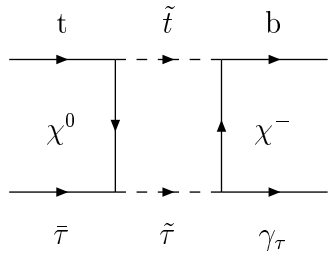
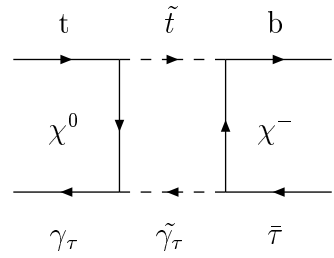
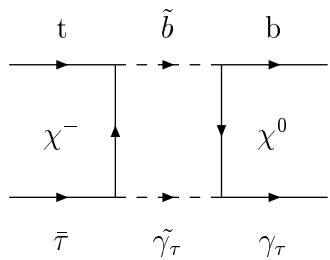
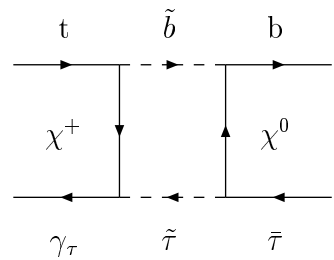
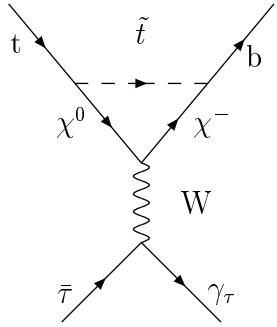
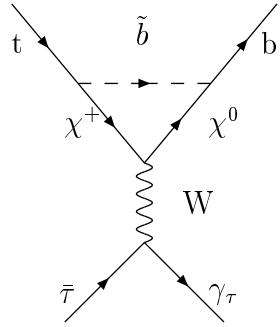
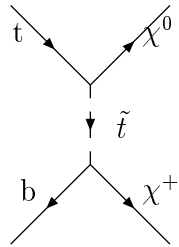
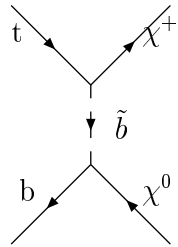
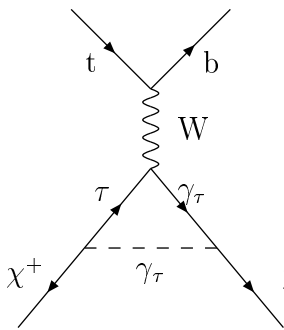
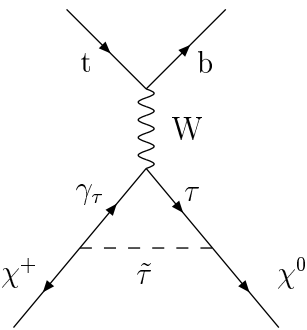
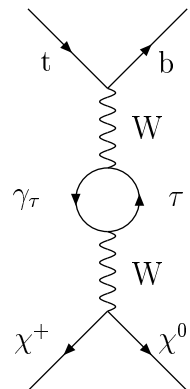
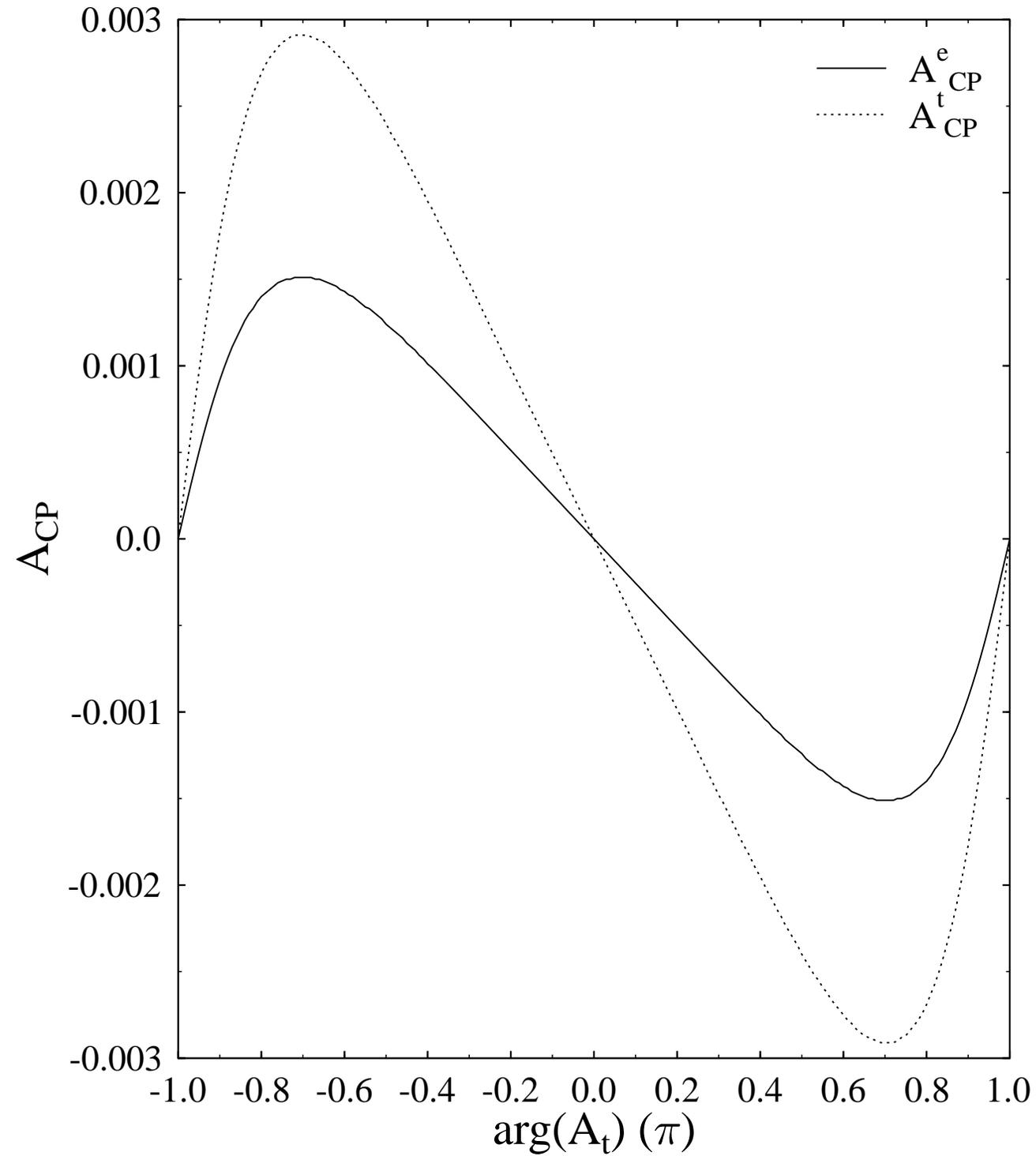
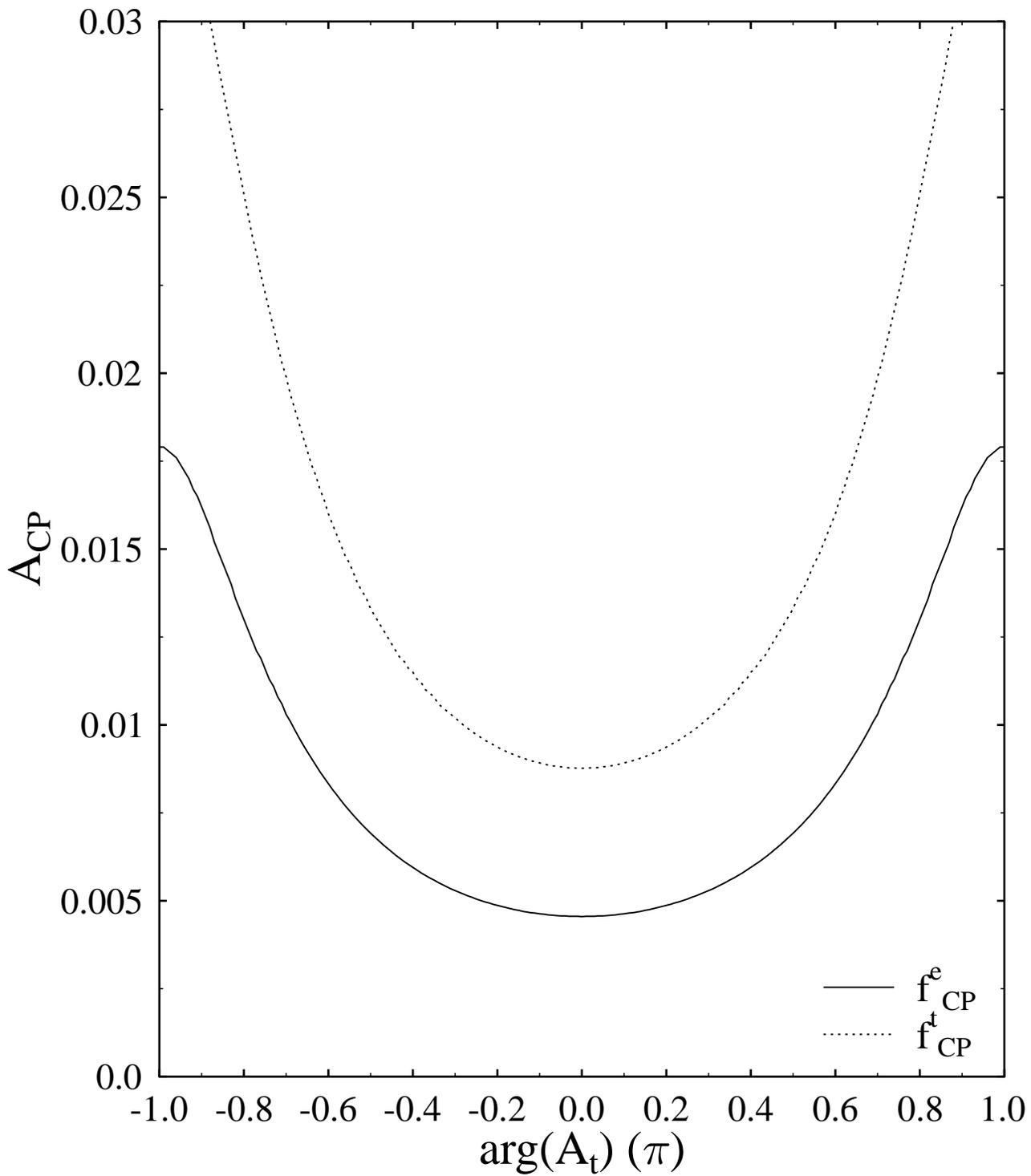
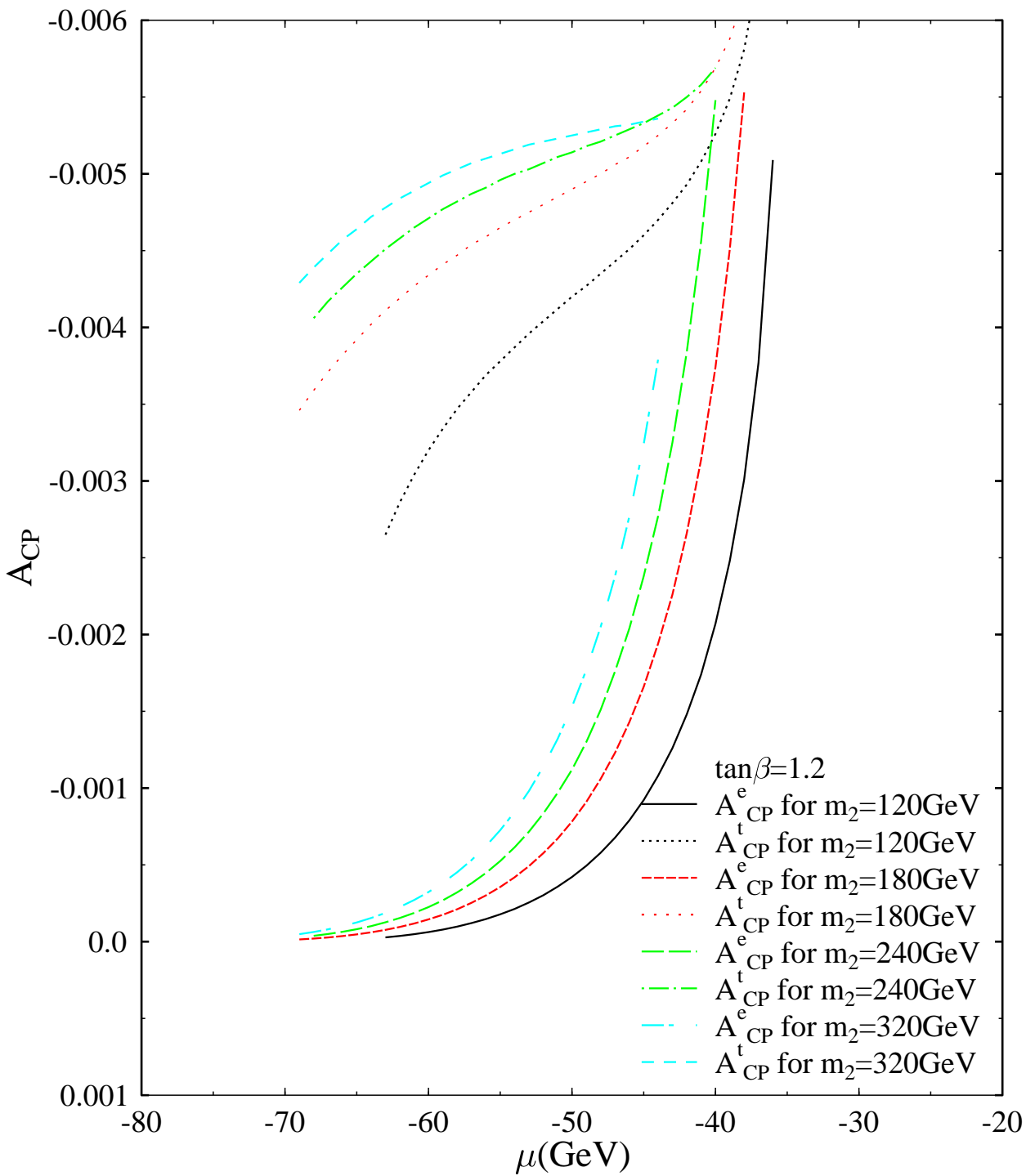
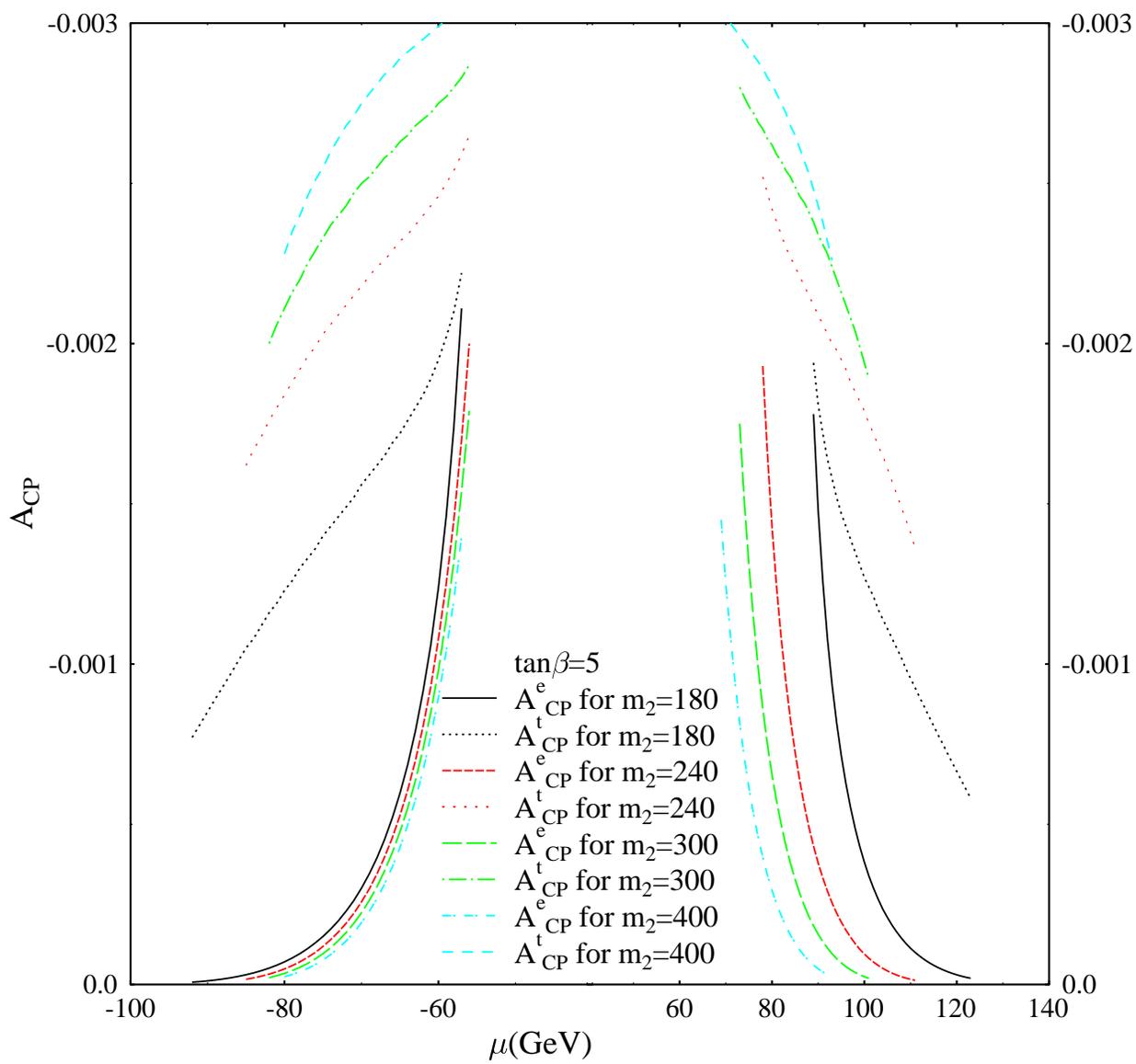
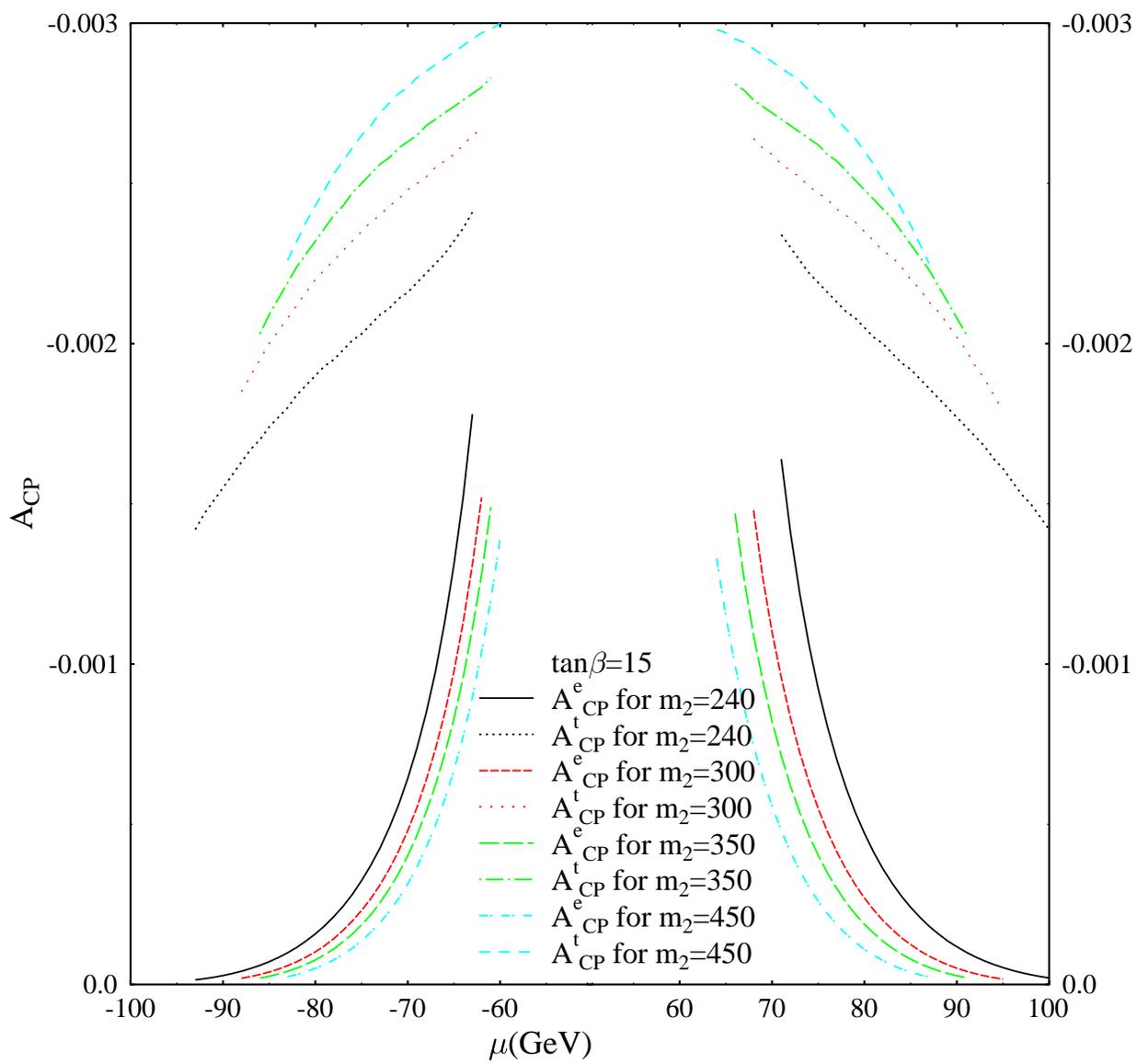
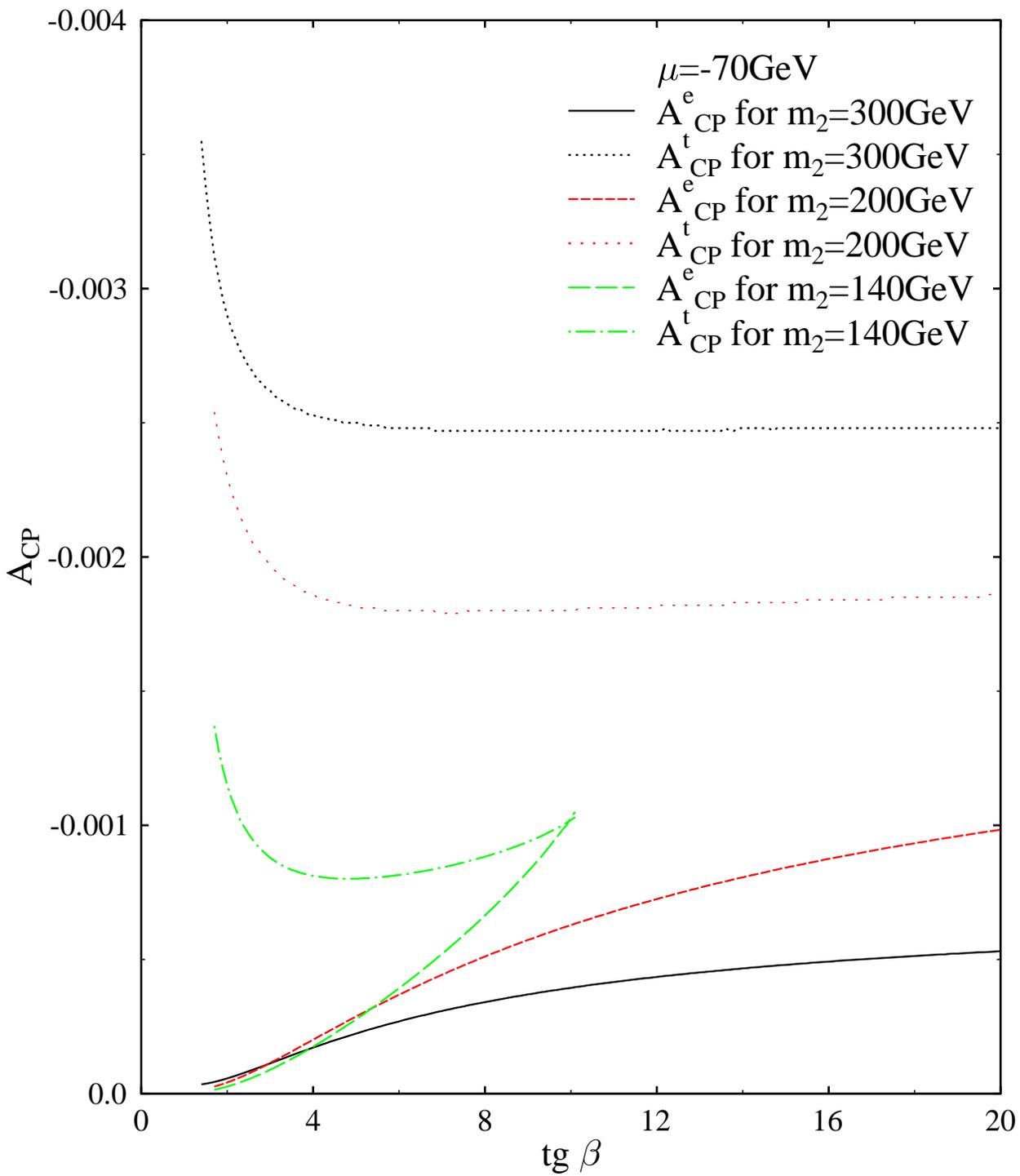
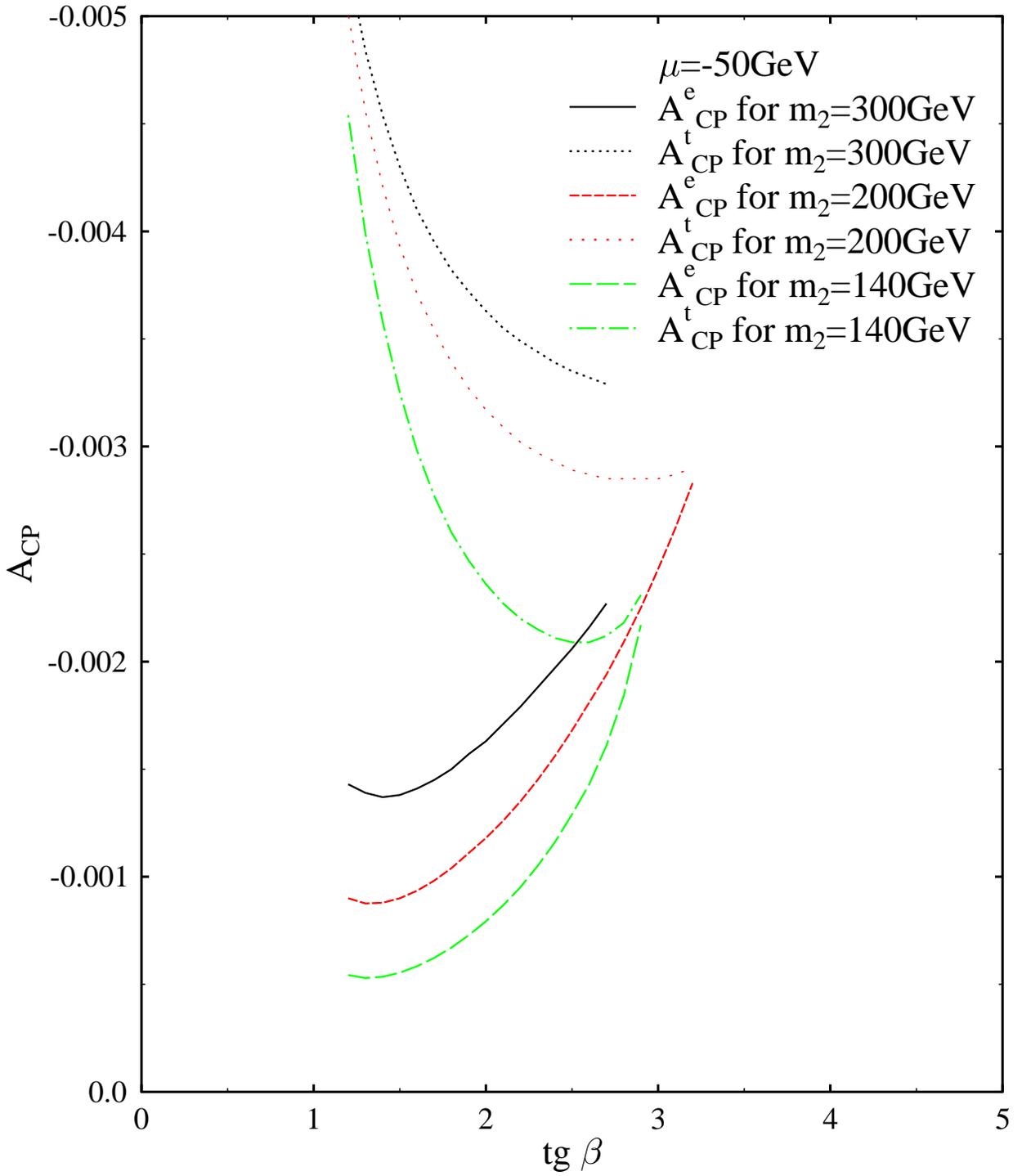
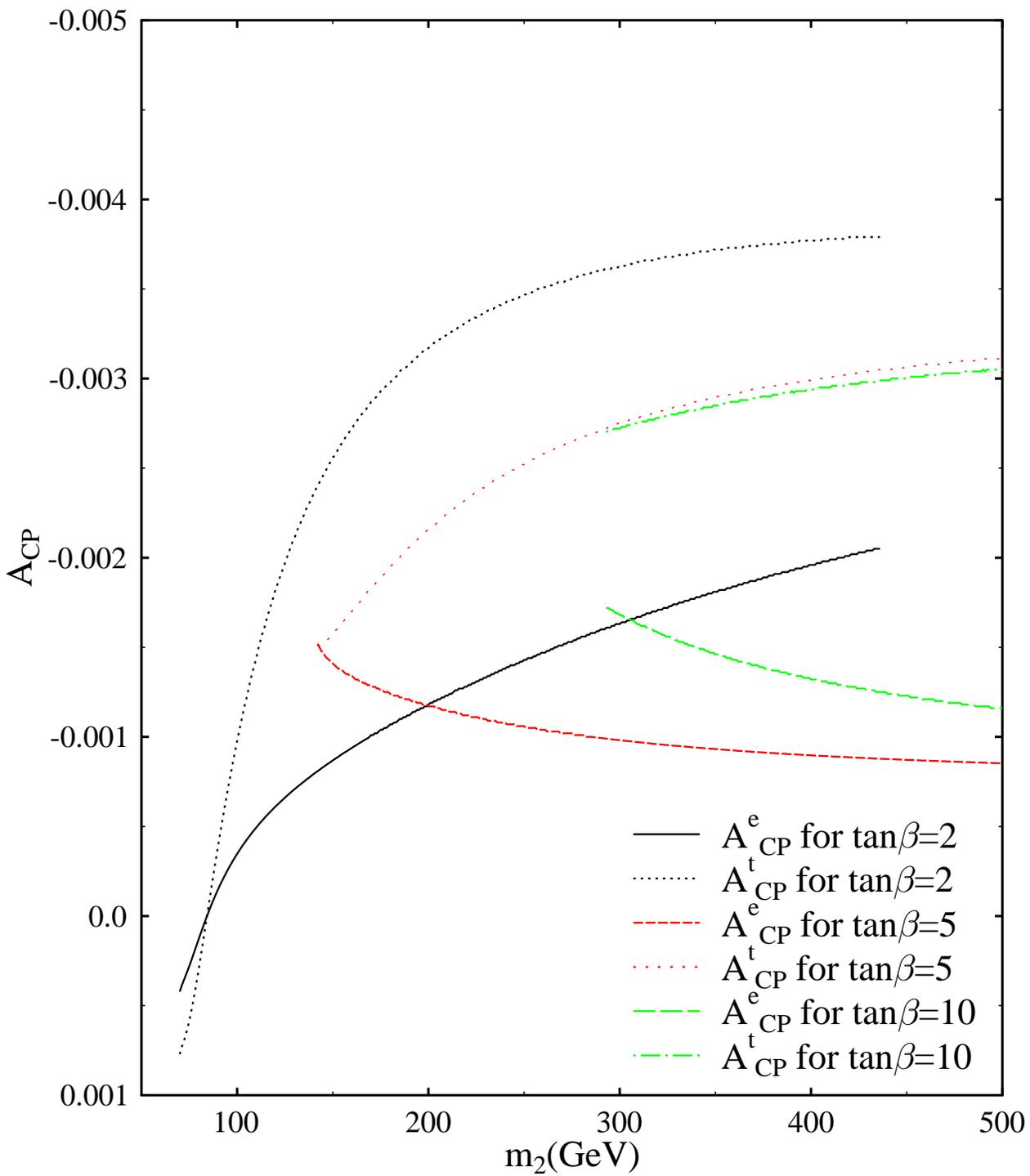
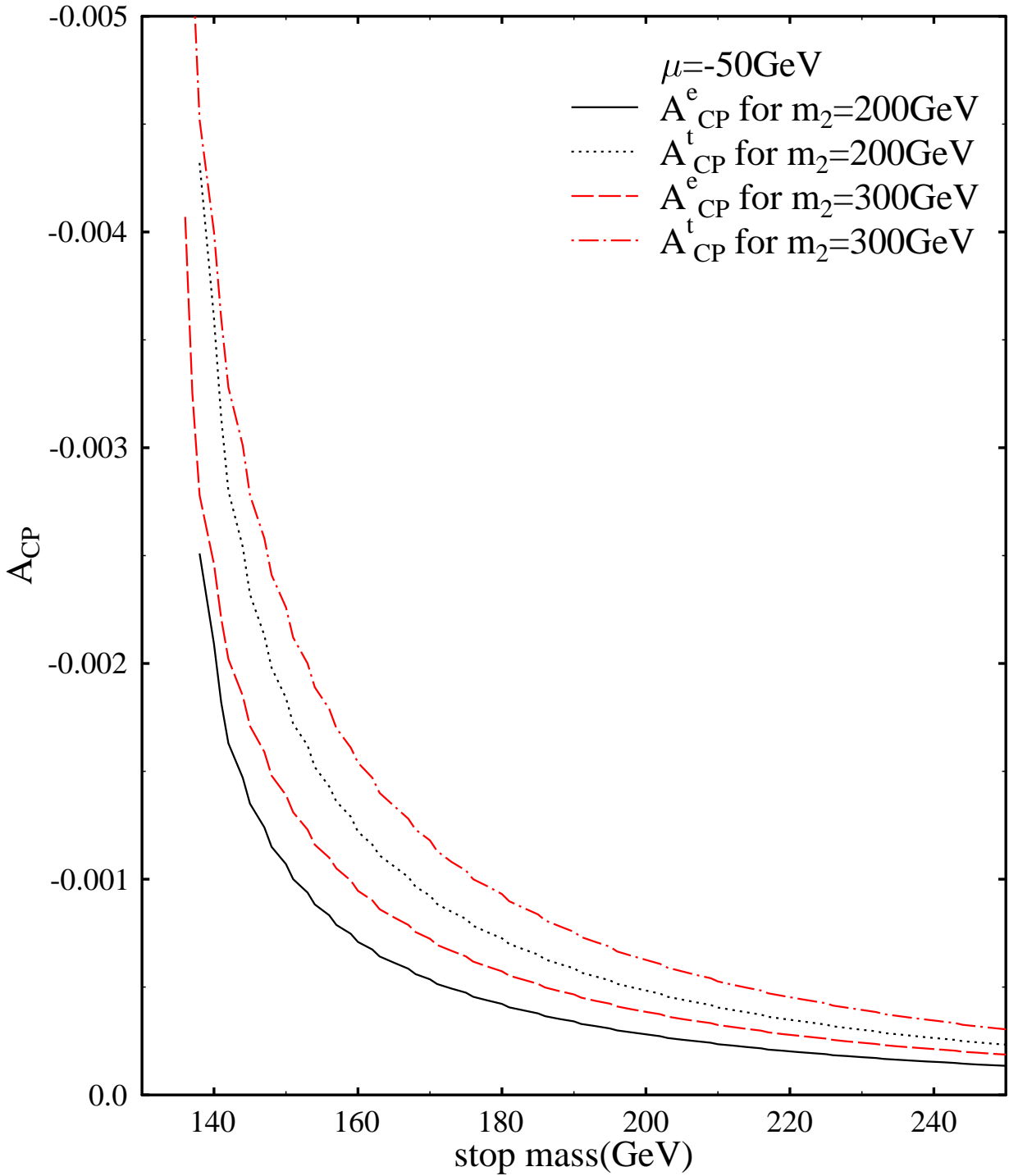
\end{document}